%% file: main.tex
\documentclass[conference,compsoc]{IEEEtran}
\usepackage{amsmath}
\usepackage{algorithmic}
\usepackage{url}
\usepackage{graphicx}
\usepackage{array}
\usepackage{calc}
\usepackage{environ}
\usepackage{tikz}
\usetikzlibrary{calc, matrix}
\usepackage{caption}
\usepackage{subcaption}

\makeatletter
\let\matamp=&
\catcode`\&=13
\makeatletter
\def&{\iftikz@is@matrix
  \pgfmatrixnextcell
  \else
  \matamp
  \fi}
\makeatother

\newcounter{lines}
\def\endlr{\stepcounter{lines}\\}

\newcounter{vtml}
\newif\ifvtimelinetitle
\newif\ifvtimebottomline
\tikzset{description/.style={
  column 2/.append style={#1}
 },
 timeline color/.store in=\vtmlcolor,
 timeline color=red!80!black,
 timeline color st/.style={fill=\vtmlcolor,draw=\vtmlcolor},
 use timeline header/.is if=vtimelinetitle,
 use timeline header=false,
 add bottom line/.is if=vtimebottomline,
 add bottom line=false,
 timeline title/.store in=\vtimelinetitle,
 timeline title={},
 line offset/.store in=\lineoffset,
 line offset=4pt,
}

\NewEnviron{vtimeline}[1][]{%
\setcounter{lines}{1}%
\stepcounter{vtml}%
\begin{tikzpicture}[column 1/.style={anchor=east},
 column 2/.style={anchor=west},
 text depth=0pt,text height=1ex,
 row sep=1ex,
 column sep=1em,
 #1
]
\matrix(vtimeline\thevtml)[matrix of nodes]{\BODY};
\pgfmathtruncatemacro\endmtx{\thelines-1}
\path[timeline color st] 
($(vtimeline\thevtml-1-1.north east)!0.5!(vtimeline\thevtml-1-2.north west)$)--
($(vtimeline\thevtml-\endmtx-1.south east)!0.5!(vtimeline\thevtml-\endmtx-2.south west)$);
\foreach \x in {1,...,\endmtx}{
 \node[circle,timeline color st, inner sep=0.15pt, draw=white, thick] 
 (vtimeline\thevtml-c-\x) at 
 ($(vtimeline\thevtml-\x-1.east)!0.5!(vtimeline\thevtml-\x-2.west)$){};
 \draw[timeline color st](vtimeline\thevtml-c-\x.west)--++(-3pt,0);
 }
 \ifvtimelinetitle%
  \draw[timeline color st]([yshift=\lineoffset]vtimeline\thevtml.north west)--
  ([yshift=\lineoffset]vtimeline\thevtml.north east);
  \node[anchor=west,yshift=16pt,font=\large]
   at (vtimeline\thevtml-1-1.north west) 
   {\textsc{Timeline \thevtml}: \textit{\vtimelinetitle}};
 \else%
  \relax%
 \fi%
 \ifvtimebottomline%
   \draw[timeline color st]([yshift=-\lineoffset]vtimeline\thevtml.south west)--
  ([yshift=-\lineoffset]vtimeline\thevtml.south east);
 \else%
   \relax%
 \fi%
\end{tikzpicture}
}

\begin{document}
\title{A Survey and Evaluation of Data Center Network Topologies}

\author{\IEEEauthorblockN{Brian Lebiednik, Aman Mangal}
\IEEEauthorblockA{\{blebiednik3, amanmangal\}@gatech.edu\\
School of Computer Science\\College of Computing\\
Georgia Institute of Technology\\
Atlanta, Georgia 30332--0250}
\and
\IEEEauthorblockN{Niharika Tiwari}
\IEEEauthorblockA{ntiwari6@gatech.edu\\
School of Electrical and Computer Engineering\\
College of Engineering\\
Georgia Institute of Technology\\
Atlanta, Georgia 30332--0250}}

\maketitle
\thispagestyle{plain}
\pagestyle{plain}

\begin{abstract}
Data centers are becoming increasingly popular for their
flexibility and processing capabilities in the modern computing environment.
They are managed by a single entity (administrator)
and allow dynamic resource provisioning, performance optimization
as well as efficient utilization of available resources.
Each data center consists of massive compute, network and
storage resources connected with physical wires. The large scale nature of data centers requires careful
planning of compute, storage, network nodes, interconnection as well as inter-communication 
for their effective and efficient operations.
In this paper, we present a comprehensive survey and taxonomy of
network topologies either used in commercial data centers,
or proposed by researchers working in this space.
We also compare and evaluate some of those topologies
using \textit{mininet} as well as \textit{gem5} simulator
for different traffic patterns,
based on various metrics including throughput, latency and bisection bandwidth.
\end{abstract}

\section{Introduction}
\input{introduction}

\section{Background}
\label{sec:background}
\input{background}

\section{Comparison with NoCs}
\label{sec:compare_nocs}
\input{noc.tex}

\section{History of Data Center Networks}
\label{sec:history}
\input{history}

\section{Taxonomy of DCN Topologies}
\label{sec:taxonomy}
\input{taxonomy}

\section{Evaluation}
\label{sec:evaluation}
\input{evaluation}

\section{Conclusion \& Future Work}
\label{sec:conclusion}
\input{conclusion}

\section*{Acknowledgments}
Many thanks to Prof Tushar Krishna for guiding us on
every step of the project, for providing us with computing
resources for our simulations and experiments,
for giving us periodic feedback
and for the amazing course offered as CS8803-ICN at
Georgia Tech, where we acquired all the background
to allow us to write this paper. Also thanks to the reviewers of our paper the provided us with several good points on where to continue our testing and improve our paper. Specifically, one reviewer asked for more information on how to we were going to conduct our testing in Mininet. Prof Krishna also asked for us to scale the number of cores which we did in Mininet.

\bibliographystyle{IEEEtran}
\bibliography{icn}
\end{document}

%% file: introduction.tex
A data center is a facility used to house
computer systems and associated components,
such as telecommunications and storage systems
\cite{wiki_data_center}.
They are key enabler for cloud computing to
provide Software-as-a-service (SaaS),
Infrastructure-as-a-service
(IaaS) for online web services, big data computing,
large simulations etc.
Today's data center network (DCN) contains thousands of compute nodes
with significant network bandwidth requirements.
Companies like Amazon, Google and Microsoft
are building large data centers for cloud computing
\cite{2009tech} to keep up with application demands.
Recent trends show companies like Dropbox, Apple embracing
the move to build their own private cloud in order to gain
better control, security and higher
efficiency \cite{2016inside_apple, 2016scaling_dropbox}.
The popularity of cloud, scale of data centers
and desire to achieve highest level of
application performance requires careful planning
of compute, storage and the interconnection network or
topology.

\begin{figure}[h]
\centering
\includegraphics[scale=0.3]{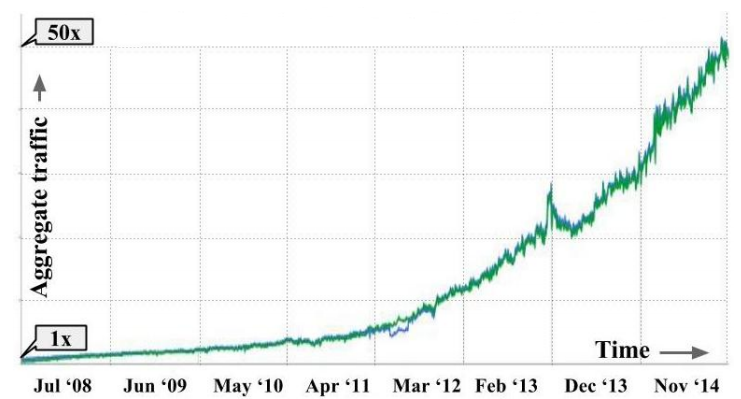}
\caption{Aggregate server traffic in Google's Data Centers
\label{fig:demands}}
\end{figure}

While data centers offer tremendous benefits,
bandwidth demands are doubling every
12-15 months as shown in Figure \ref{fig:demands}.
A number of recent trends drive this growth.
Many data center applications require bandwidth intensive
one-to-one, one-to-several
(e.g. distributed file systems \cite{ghemawat2003google}),
one-to-all (e.g. application data broadcasting),
or all-to-all (e.g. MapReduce) \cite{dean2008mapreduce}
 communication.
Data set sizes are continuing to explode with more
photo/video content, logs, and the proliferation of
internet-connected sensors. As a result,
network intensive data processing pipelines must
operate over ever-larger data sets.
However, today, deployment of even the highest-end enterprise
network equipment only delivers 50\% of the available bandwidth.
The vital challenges faced by the legacy
DCN architecture trigger the need for new DCN architectures,
to accommodate the growing demands of the cloud computing paradigm.

In this paper, we present history and taxonomy of
various DCN topologies that have been proposed so far
and how they have advanced the state-of-the-art technology
to overcome aforementioned challenges.
The main focus while designing the DCN architecture,
has been scalability, cost, latency, extensiblity.
Further, we have implemented Google Fat Tree
\cite{al2008scalable}, Facebook Fat Tree \cite{facebook2014}
and DCell \cite{guo2008dcell} on two different network
simulators - \textit{gem5} \cite{binkert2011gem5}
and \textit{mininet} \cite{de2014using}.
We present our evaluation results and compare latency
and throughput metrics for different network traffic patterns.
With this work, we hope to present a general
overview of various DCN topologies,
as well as experimental results to corroborate our analysis.

\begin{figure*}[ht]
\centering
\includegraphics[scale=0.9]{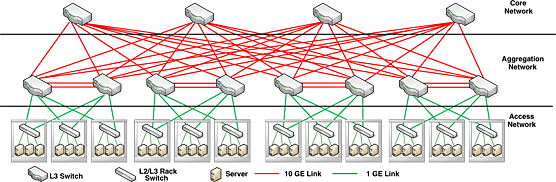}
\caption{Legacy DCN Architecture
\label{fig:dcn_architecture}}
\end{figure*}

This paper is organized as follows. Section \ref{sec:background}
elaborates on general architecture of data centers along with
illustrating terms that are used frequently in this space.
Section \ref{sec:compare_nocs} explains fundamental differences
in topologies for DCN compared to Network on Chip (NoCs).
We provide summary of various topologies that have been used
in commercial data centers or proposed by researchers
in Section \ref{sec:history}.
Taxonomy of DCN topologies is presented in Section
\ref{sec:taxonomy}. Section \ref{sec:evaluation} presents
simulation results from our experiments on both \textit{gem5}
and \textit{mininet}.
Finally, we discuss our conclusion and future work in Section
\ref{sec:conclusion}.

%% file: background.tex
Data Center Networks (DCN) today, are typically based on the three
tier, hierarchical, tree based architecture as shown in figure
\ref{fig:dcn_architecture}.
It has a core tier at the root of the tree, an aggregation
layer in the middle and edge tier at the leaves of tree.
A group of hosts (mostly 16/32) is connected to one switch,
called ToR switch (Top of Rack), building the edge (access) layer
of the hierarchical tree structure.
Core layer and aggregation layer uses high end switches
aggregating the traffic coming from the lower layers.
Routing is performed by traversing up the tree until the
lowest common ancestor and then, down the tree to reach
to the final destination.
There exists redundant paths among two hosts in the
network allowing packet delivery in case of switch failures.
Unlike Network on Chip (NoCs), data center hosts and switches
implement much more complex protocols to ensure reliability
of communication.
We discuss some of the most widely used protocol as follows.

\subsection{Ethernet}
Ethernet is a data link layer protocol to send packets (frames)
from one point to another point (host or switch),
directly connected to each other.
It provides best effort service based on collision detection (CD)
without any flow control.
Frames are dropped if the queues are full at the receiver
without notifying the sender.
Further, Ethernet switches are similar to a NoC crossbar
and implement FIFO model for packet processing.

\subsection{TCP/IP}
IP is a network layer protocol to ensure routing of packets
from one host to any other host in the network.
TCP runs on top of IP layer and implements: (a) flow control to
prevent receiver's buffer from overflowing, (b) re-transmission
to ensure reliable data transfer, and (c) congestion control to
minimize packet loss. Note that data centers, therefore, only
implement end to end flow control and there is no mechanism
to ensure point to point (i.e. link level) packet delivery
as Ethernet only provides best effort service.

\subsection{Over-subscription}
Many data center designs introduce over-subscription as a means
to lower the total cost of the design. We define the term
over-subscription to be the ratio of the worst-case achievable
aggregate bandwidth among the end hosts to the total bisection
bandwidth of a particular communication topology.
An over-subscription of 1:1 indicates that all hosts may
potentially communicate with arbitrary
other hosts at the full bandwidth of their network interface.
An over-subscription value
of 5:1 means that only 20\% of available host bandwidth is available
for some communication patterns.
Although data centers with over-subscription of 1:1 are possible,
the cost for such designs is typically prohibitive, even for
modest-size data centers \cite{al2008scalable}.

%% file: noc.tex
Data Center Network topologies are inspired from the world
of Network on Chip (NoC). However, there are some
key differences. We highlight some of them as below -
\begin{itemize}
\item \textbf{\textit{High Radix Routers:} }
Data centers typically employ high radix routers in order to
utilize more path diversity to achieve higher throughput.
The number of links is not a concern in comparison to
on chip networks, as space is an inexpensive commodity.

\item \textbf{\textit{Link Bandwidths:} }
In off chip networks, the bandwidth available
usually differ among links belonging to different levels (tiers)
in the hierarchy. This allows the number of downlinks at a router
to exceed the number of uplinks.
Hence, to equalize total incoming and outgoing bandwidth,
link capacity is usually seen to increase
as we go up the hierarchy.

\item \textbf{\textit{Routing Algorithms: }}
NoCs commonly tend to be two/three dimensional
and hence, adopt dimensional routing such as XY routing,
turn model based routing. Whereas, off chip networks tend
to use algorithms such as ECMP (Equal Cost Multipath
Routing) to make use of redundant paths in the network.

\item \textbf{\textit{Flow Control: }}
Off chips networks do not perform any kind of link level
flow control unlike NoCs as discussed in section \ref{sec:background}.
A packet may get dropped when output output buffers are full.

\item \textbf{\textit{Routing delay and Link Latency: }}
Both routing delay and link latency tend to be typically
higher in off chip networks, merely due to much larger
size of the components. That is why, diameter for off
chip topology is typically smaller compared to NoCs.
\end{itemize}

%% file: history.tex
\begin{figure}[h]
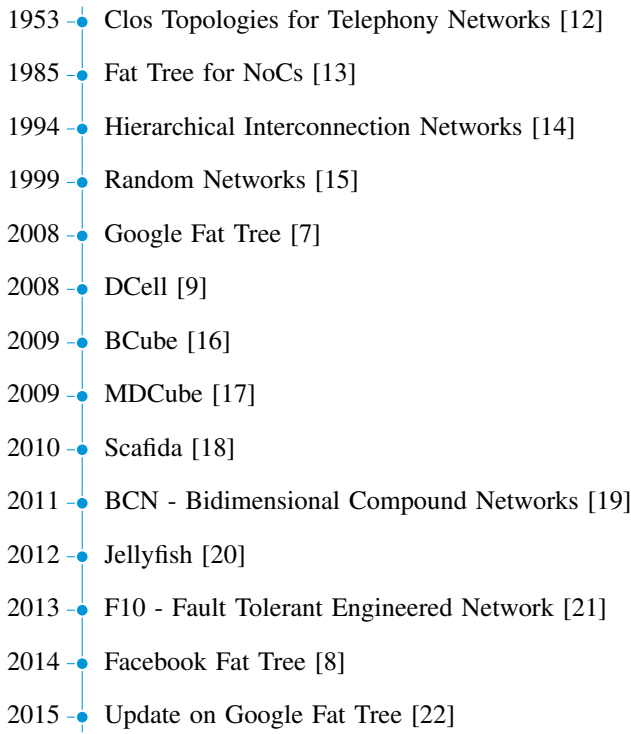

\centering
\begin{vtimeline}[timeline color=cyan!80!blue,
add bottom line,
line offset=2pt,
description={text width=7.2cm},
row sep=2ex]
1953 & Clos Topologies for Telephony Networks \cite{clos1953study} \endlr
1985 & Fat Tree for NoCs \cite{leiserson1985fat} \endlr
1994 & Hierarchical Interconnection Networks \cite{breznay1994class} \endlr
1999 & Random Networks \cite{barabasi1999emergence} \endlr
2008 & Google Fat Tree \cite{al2008scalable} \endlr
2008 & DCell \cite{guo2008dcell} \endlr
2009 & BCube \cite{guo2009bcube} \endlr
2009 & MDCube \cite{wu2009mdcube} \endlr
2010 & Scafida \cite{gyarmati2010scafida} \endlr
2011 & BCN - Bidimensional Compound Networks \cite{guo2011bcn} \endlr
2012 & Jellyfish \cite{singla2012jellyfish} \endlr
2013 & F10 - Fault Tolerant Engineered Network \cite{liu2013f10} \endlr
2014 & Facebook Fat Tree \cite{facebook2014} \endlr
2015 & Update on Google Fat Tree \cite{singh2015jupiter} \endlr
\end{vtimeline}
\caption{Timeline of DCN Topologies \label{fig:timeline}}
\end{figure}

In this section, we describe some of the data center network
(DCN) topologies that have been proposed over the time.
We have organized all of these topologies on a timeline
based on when the corresponding paper was published,
as shown in figure \ref{fig:timeline}.

\subsection{Fat Tree Topology}
\begin{figure}[h]
\centering
\includegraphics[scale=0.6]{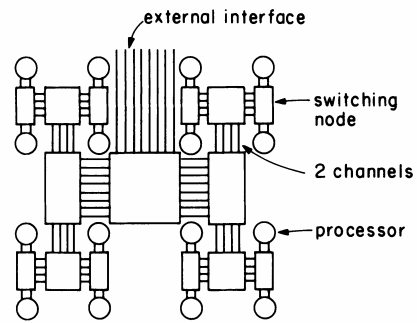}
\caption{Fat Tree for NoCs (taken from \cite{leiserson1985fat})\label{fig:fat_tree}}
\end{figure}

More than 50 years ago, Charles Clos proposed non-blocking
\textit{Clos} network topologies for telephone circuits
\cite{clos1953study} that delivers high bandwidth.
Many of the commercial data center networks adopt a special
instance of \textit{clos} topologies called Fat Tree
\cite{leiserson1985fat}.
FT was originally proposed for on chip networks (NoCs)
organizing processors in a complete binary tree
as show in Figure \ref{fig:fat_tree}.
Each Processor is connected to one router (switching node)
with a duplex link (two channels/links - one uplink and
other downlink).
Packet routing is also highly simplified and requires only
$2 \log(n)$ space for destination.
Any node can be reached from any other node by traversing
a unique path through the common ancestor.
Fat Tree topologies are popular for their non-blocking
nature, providing many redundant paths between any 2 hosts.
Such topologies are later used to build fast and efficient
super computers such as BlackWidow \cite{scott2006blackwidow}
along with successful use in commercial data centers
\cite{al2008scalable, facebook2014}.

\subsection{Google Fat Tree}
\begin{figure}[h]
\centering
\includegraphics[scale=0.4]{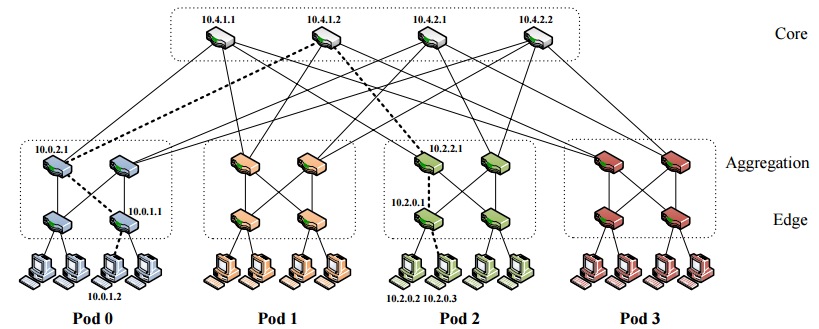}
\caption{Google Fat Tree Topology (taken from \cite{al2008scalable})\label{fig:gfat_tree}}
\end{figure}

Google implemented a slight modification of Fat Tree topology
to interconnect commodity Ethernet switches in order to
produce scalable large data centers \cite{al2008scalable}.
The topology consists of k-port routers along with commodity
compute nodes at the leaves of the tree as shown in
figure \ref{fig:gfat_tree}.
The basic building block of the data center is called a pod.
A Fat Tree consists of k pods, each containing two layers of k/2
switches. Each k-port switch in the lower layer is directly
connected to k/2 hosts. Each of the remaining k/2 ports
is connected to k/2 of the k ports in the aggregation
layer of the hierarchy. There are $(k/2)^2$ k-port core switches.
Each core switch has one port connected to each of the k pods.

Routing in Fat Tree is implemented as follows.
Each flowlet takes a unique path out of all possible
paths to reach the common ancestor while avoiding
reordering of packets. A flowlet is a collection
of TCP segments (packets) sent in a quick succession.
After reaching the common ancestor, it traverses
downwards taking the only possible path.
This design allows to build the data centers using
commodity switches instead of very expensive routers
reducing the overall cost significantly.
It can use all the available redundant paths to send packets
between two nodes while also benefiting from adaptive routing.

\subsection{DCell}
\begin{figure}[h]
\centering
\includegraphics[scale=0.3]{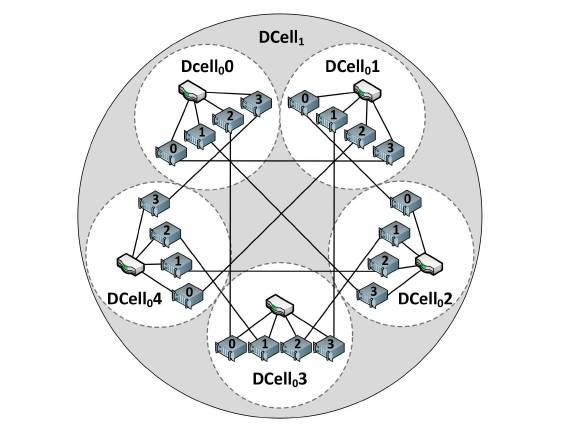}
\caption{DCell1 network when n=4, composed of 5 DCell0,
forming a fully connected graph.
(taken from \cite{guo2008dcell})\label{fig:dcell}}
\end{figure}

DCell is a server-centric hybrid DCN architecture
where one server is directly connected to many other
servers \cite{guo2008dcell}.
A server in a DCell is equipped with multiple Network interface
cards (NICs). The DCell follows a recursively build hierarchy
of cells as shown in Figure \ref{fig:dcell}.
A cell0 is the basic unit and building block of DCell
topology arranged in multiple levels, where a higher level cell
contains multiple lower layer cells.
A cell0 contains n servers and one commodity 
network switch.
The network switch is only used to connect the server
within a cell0. A cell1 contains $k=n+1$ cell0 cells,
and similarly a cell2 contains $k*n+1$ cell1.
A Dcell can be built recursively resulting
in more than 3.26 million servers with an average
diameter of less than 10 (k=3, n=6).
Routing in Dcell follows a divide and conquer approach.
For packets to reach from a source host to destination host,
it needs to traverse from source to common ancestor DCell,
a link connecting the previous level DCells and
finally, to the destination.
The exact path can be found similarly in a recursive fashion.
The protocol is further extended to implement
fault tolerant routing (DFR)
to cope with link or node failures. Overall,
DCell is highly scalable and fault tolerant topology
however, it provides low bisection bandwidth.

\subsection{BCube}
\begin{figure}[h]
\centering
\includegraphics[scale=0.35]{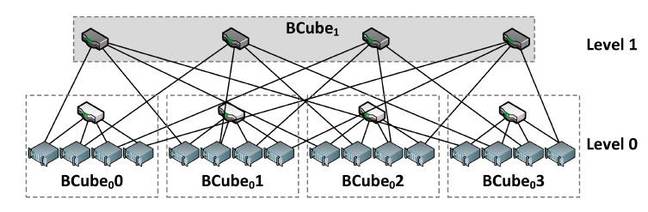}
\caption{BCube Topology (taken from \cite{guo2009bcube})\label{fig:bcube}}
\end{figure}

BCube network architecture takes a server-centric approach to
produce a modular data center (MDC) using commodity
switches. It places intelligence on
MDC servers and works with low end COTS mini switches.
There are two types of devices in BCube: servers with
multiple ports and switches that connect to a constant number
of servers. It is recursively defined structure with $BCube_0$
simply being n servers connected to an n-port switch.
$BCube_k$ is constructed with n $BCube_{k-1}$ having
$n^k - 1$ switches each connecting same index server
from all the $BCube_{k-1}$.
With 8-port mini-switches,
it can support up to 4096 servers in one $BCube_3$.
The figure \ref{fig:bcube} shows a $BCube_{1}$
with n=4 with 2 levels. Source based routing is 
performed using intermediate nodes as packet forwarder
ensuring, decreasing hamming distance between each
consecutive intermediate host to the destination.
Periodic search for optimal path is performed in order
to cope with failures in the network.
One-to-all, all-to-one and all-to-all traffic can also
be routed by using redundant (k+1) ports at each hosts.

\subsection{MDCube}
\begin{figure}[h]
\centering
\includegraphics[scale=0.33]{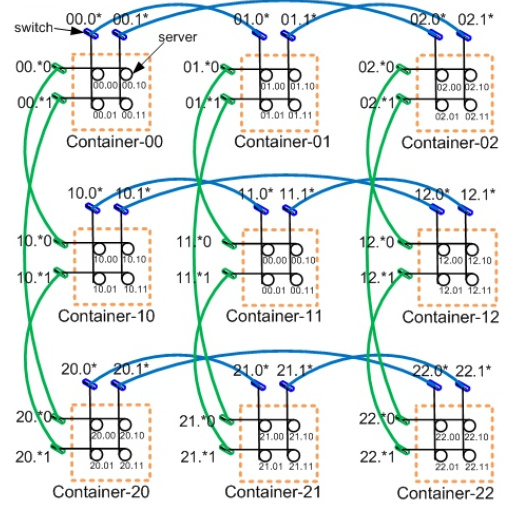}
\caption{A 2-D MDCube Topology constructed from 9=3*3
$BCube_1$ containers with n=2, k=1 (taken from \cite{wu2009mdcube})\label{fig:mdcube}}
\end{figure}

In order to build a large data center, MDCube \cite{wu2009mdcube}
uses BCube \cite{guo2009bcube} as its building block
and allows to interconnect hundreds and thousands of
BCube containers in 1-D or 2-D fashion to achieve high
network capacity as shown in figure \ref{fig:mdcube}.
It connects two containers in the same dimension (i.e. row or column)
with a direct link to form a basic complete graph among all
containers similar to a Flattened Butterfly.
Single path routing is performed
topology by finding a pair of switches in
an intermediate container.

\subsection{Scafida}
\begin{figure}[h]
\centering
\begin{subfigure}[b]{.23\textwidth}
  \centering
  \includegraphics[scale=0.25]{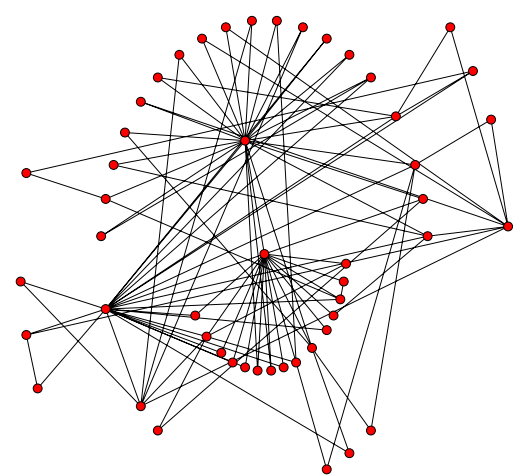}
  \caption{}
\end{subfigure}
\begin{subfigure}[b]{.23\textwidth}
  \centering
  \includegraphics[scale=0.25]{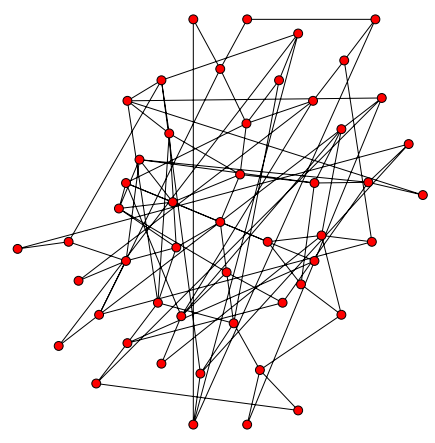}
  \caption{}
\end{subfigure}
\caption{(a) Scale Free Network (SFN), (b) SFN with maximal degree 5
(taken from \cite{gyarmati2010scafida}) \label{fig:scafida}}
\end{figure}

Scafida \cite{gyarmati2010scafida} is a asymmetric scale-free
data center network topology
to achieve short distance, high error tolerance and incremental
build. Scale-free networks have two important properties - small
diameter and high resistance to random failures.
The same set of properties are highly desirable in data
center network topologies.
Scafida provides methodologies to construct such a topology
for data centers while making reasonable modifications to
original scale-free network paradigm \cite{barabasi1999emergence}.
Scafida consists of heterogeneous set of switches and hosts in
terms number of ports/links/interfaces. The topology is built
incrementally by adding a node and then, randomly connecting all the
available ports to existing empty ports.
The number of ports are limited by the available ports on a node
unlike original scale-free networks.
Such a network provides high fault tolerance. Results show that
even if 20\% of the switches fail, more than 90\%
of the server pairs still have 2 disjoint paths.
Examples of a scale-free topology is shown in figure \ref{fig:scafida}.
No routing algorithm is proposed yet for such networks though,
the idea of random construction of a data center looks promising.
However, wiring, handling failure of nodes with large degree,
routing algorithm are still major issues that needs to be addressed.

\subsection{HCN \& BCN}
\begin{figure}[h]
\centering
\includegraphics[scale=0.35]{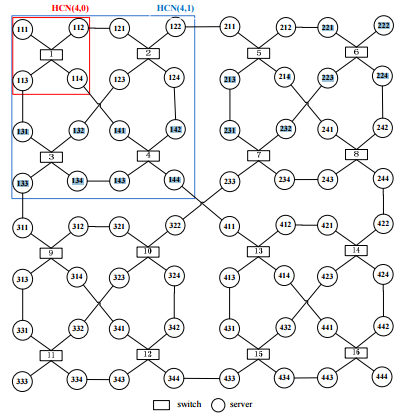}
\caption{HCN(n=4, h=2)
(taken from \cite{guo2011bcn}) \label{fig:bcn}}
\end{figure}

Hierarchical Irregular Compound Network (HCN) and Bidimensional
Compound Network (BCN) \cite{guo2011bcn}
are dual-port-server based, symmetric, regular
and extensible architectures. HCN is recursively defined structure.
$HCN(n, 0)$ is the base case, consisting of n dual port servers,
each of them connected to n port switch on one of the two ports.
Each of the server will have 1 port free, resulting in $n$
total free ports in $HCN(n, 0)$.
A $HCN(n, 1)$ is then, constructed using $n$ $HCN(n, 0)$ modules
by connecting $(n-1)$ out of $n$ available ports of each $HCN(n, 0)$
with rest of the $(n-1)$ $HCN(n, 0)$.
Each of the $HCN(n, 0)$ module will have 1 port left free and
allowing a further extension using total of n free ports in
$HCN(n, 1)$. In general, a high-level $HCN(n, h)$
employs n modules of $HCN(n, h-1)$
consisting of n free ports for further extending the topology.
An example of $HCN(4, 2)$ is shown in figure \ref{fig:bcn}.

BCN is multi-level irregular compound graph recursively
defined in the first dimension, and a level one regular compound
graph in the second dimension. $BCN(\alpha, \beta, 0)$
has $(n=\alpha + \beta)$ where only $\alpha$ ports are available
for further extension similar to HCN. These $\alpha$ servers having the
$\alpha$ ports available are called master servers.
In the second dimension, it is a closed structure (i.e.
cannot be extended further) and constructs a fully connected
graph of $BCN(\alpha, \beta, h)$ modules on the available
$\alpha^h\cdot\beta$ servers, called slave servers.
This provides flexibility in extensiblility of the topology
as necessary by controlling the parameters $\alpha$ and $\beta$.
Routing in BCN is performed similar by recursively finding
an intermediate link that interconnects the two
BCN modules where source and destination are located.

\subsection{Jellyfish}
Jellyfish is a flexibility and high bandwidth oriented
network topology consisting of n port switches. Each switch has
$r$ ports connected to other switches and rest of the $k=(n-r)$ ports
connected to hosts. The links are added by randomly connecting a
pair of switches that are not already connected (i.e. not neighbors)
and having at least one free port. The topology can be further
extended by removing existing $(x, y)$ link and adding
$(x, p_1)$ and $(x, p_2$) link where $p_1$, $p_2$ are free ports
on the new switch. Such random graphs have higher throughput
because they have low average path lengths in comparison to symmetric
topologies such as fat tree. However, routing, packaging issues needs
to be addressed for practical use of the topology.

\subsection{F10 (Fault Tolerant Engineered Network)}
\begin{figure}[h]
\centering
\includegraphics[scale=0.35]{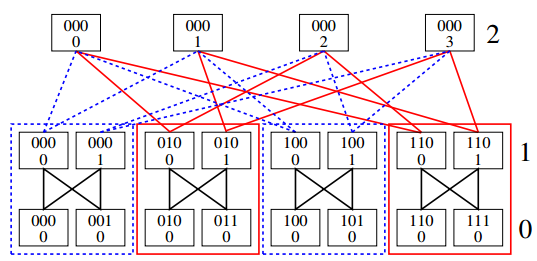}
\caption{F10 Topology, blue links are part of A sub-tree,
red links are part of B sub-tree
(taken from \cite{liu2013f10}) \label{fig:ften}}
\end{figure}

F10 \cite{liu2013f10} is a simple modification to Fat Tree
topology to gain better fault tolerance properties.
The key weakness in the standard Fat Tree is that all
sub-trees at level $i$ are wired to the parents at level $i+1$ in
an identical fashion. A parent attempting to detour around
a failed child must use roundabout paths (with inflation
of at least four hops) because all paths from its rest of
the children to the target sub-tree use the same failed
node. The AB FatTree in F10 solves this problem by defining
two types of sub-trees (called type A and type B) that are
wired to their parents in two different ways
as show in figure \ref{fig:ften}. With this
simple change, a parent with a failed child in a type A
sub-tree can detour to that sub-tree in two hops through the
parents of a child in a type B sub-tree (and vice versa),
because those parents do not rely on the failed node.

\subsection{Facebook Fat Tree}
\begin{figure}[h]
\centering
\includegraphics[scale=0.9]{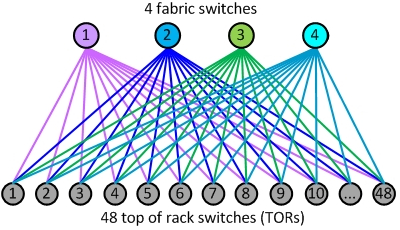}
\caption{Facebook Fat Tree Topology
(taken from \cite{facebook2014}) \label{fig:facebook}}
\end{figure}

Facebook deployed a version of Fat Tree topology in order to
achieve high bisection bandwidth, rapid network deployment and
performance scalability to keep up with the agile nature of
applications running in the data centers.
It consists of pods, a standard unit of network as show in
figure \ref{fig:facebook}. The uplink bandwidth of each TOR is
4 times (4*40G = 16*10G) the downlink bandwidth for each
server connected to it. To implement building-wide connectivity,
it created four independent "planes" of spine switches [Tier 3 switch],
each scalable up to 48 independent devices within a plane.
Border Gateway Protocol (BGP4) is used as a control
protocol for routing whereas a centralized
controller is deployed to be able to override any routing paths
whenever required, taking a "distributed control, centralized override"
approach. In order to use all the available paths between 2
hosts, ECMP (Equal Cost Multiple Path)
routing with flow based hashing is implemented.

%% file: taxonomy.tex
\newcolumntype{P}[1]{>{\centering\arraybackslash}m{#1}}

In this section, we present taxonomy of the DCN topologies
that we have discussed so far.
We believe that following criterion are subset of degrees of
freedom that are available to a data center network architecture
while designing a new topology or choosing from existing topologies.

\subsection{Build Approach}
Data centers can be built either by adding links between 2 nodes
randomly or in deterministic pattern (see Table
\ref{tab:build_approach}). Random topologies have
lower diameter but suffer from complex
routing algorithms and wiring issues.
\begin{table}[h]
\centering
\begin{tabular}{|P{1.5cm}|P{3.6cm}|P{2cm}|}
\hline
\textbf{Build Approach} & \textbf{Description} & \textbf{Examples} \\ \hline
\textbf{Random} & Add links between nodes using a randomized algorithm & Scafida, Jellyfish \\ \hline
\textbf{Deterministic} & Add links between nodes in a deterministic pattern & Fat Tree, DCell, BCN, BCube \\ \hline
\end{tabular}
\caption{DCN Topologies based on build approach}
\label{tab:build_approach}
\end{table}

\subsection{Server-Centric v/s Switch-Centric}
Some DCN topologies have hosts that take part in routing
and forwarding of packets. It requires them to have additional logic
of forwarding and routing traffic. Software routing may have
degraded performance and can affect the applications running
on the hosts.
\begin{table}[h]
\centering
\begin{tabular}{|P{1.2cm}|P{3.5cm}|P{2cm}|}
\hline
& \textbf{Description} & \textbf{Examples} \\ \hline
\textbf{Server-Centric} & Both routers and hosts forward traffic & DCell, MDCube, Scafida, BCN \\ \hline
\textbf{Switch-Centric} & Only routers forward traffic & Fat Tree, Jellyfish \\ \hline
\end{tabular}
\caption{Server v/s Switch Centric DCN Topologies}
\label{tab:centric}
\end{table}

\subsection{Direct v/s Indirect}
\begin{table}[h]
\centering
\begin{tabular}{|P{1.2cm}|P{3.5cm}|P{2cm}|}
\hline
& \textbf{Description} & \textbf{Examples} \\ \hline
\textbf{Direct} & All routers have host(s) attached to them & DCell, MDCube, Scafida, BCN \\ \hline
\textbf{Indirect} & Routers may not have host(s) attached to them & Fat Tree \\ \hline
\end{tabular}
\caption{Direct v/s Indirect DCN Topologies}
\label{tab:direct}
\end{table}

\subsection{Symmetric v/s Asymmetric}
Symmetric DCN architectures allow uniform packaging and simplified
wiring of the topology. Examples of symmetric topologies are
Fat Tree, MDCube, HCN whereas DCell, Scafida and Jellfish
are asymmetric architectures. Note that deterministic
topologies may not necessarily be symmetric such as DCell.

\subsection{Extensibility}
HCN, BCN, Scafida, Jellyfish are a few examples of extensible
DCN architectures. The size of these topologies can be easily increased
without any upper limit. However, size of Fat Tree and DCell topologies
is limited due to limited number of available ports on the switches.

\subsection{Deployment Methodology}
Data centers are built using switches and hosts.
A modular data center system is a portable method of deploying
data center capacity.
\begin{table}[h]
\centering
\begin{tabular}{|P{1.7cm}|P{2.8cm}|P{2.2cm}|}
\hline
\textbf{Deployment Methodology} & \textbf{Description} & \textbf{Examples} \\ \hline
\textbf{Modular} & Portable building block (shipping container) & BCube, MDCube \\ \hline
\textbf{Non-Modular} & Switch and host (native) & Fat Tree, DCell, Jellyfish, Scfida \\ \hline
\end{tabular}
\caption{Topologies based on deployment methodology}
\label{tab:deployment_method}
\end{table}

\subsection{Over-subscription}
Typically, DCN topologies are over-subscribed in order to reduce the
total cost of design as discussed in section \ref{sec:background}.
\begin{table}[h]
\centering
\begin{tabular}{|P{1.3cm}|P{3cm}|P{2.7cm}|}
\hline
\textbf{Over-subscription} & \textbf{Description} & \textbf{Examples} \\ \hline
\textbf{Non-blocking} & no over-subscription (over-subscription $=$ 1) & Fat Tree \\ \hline
\textbf{Blocking} & over-subscription $<$ 1 & DCell, MDCube, BCN, Scafida, Jellyfish \\ \hline
\end{tabular}
\caption{Over-subscription based DCN Topologies}
\label{tab:oversubscription}
\end{table}

\subsection{Number of Tiers (Levels)}
DCN architectures may be defined recursively in which case,
the number of levels are not preset and increases as the
size of topology increases. Further details are available in Table
\ref{tab:tiers}.
\begin{table}[h]
\centering
\begin{tabular}{|P{1.2cm}|P{3.6cm}|P{2cm}|}
\hline
\textbf{Number of Tiers} & \textbf{Description} & \textbf{Examples} \\ \hline
\textbf{Flat} & Only single tier topology & Scafida, Jellyfish \\ \hline
\textbf{Fixed} & Predefined number of tiers & Fat Tree \\ \hline
\textbf{n-tier} & Number of tiers vary as size of topology increases & DCell, BCN, MDCube \\ \hline
\end{tabular}
\caption{Number of Tiers based DCN Topologies}
\label{tab:tiers}
\end{table}

%% file: evaluation.tex
In this section, we present an experimental comparison of
Google Fat Tree, Facebook and DCell topologies using \textit{gem5}
and \textit{mininet}.

\subsection{Topologies Evaluated}
\subsubsection*{Google Fat Tree}
We implement the Google Fat Tree Topology for $(k=4)$.
It consists of 8 servers (hosts) and 20 routers (switches)
as follows: 4 core, 8 aggregate, 8 edge switches.
Each edge switch could support multiple hosts for further testing.
The limiting factor for this topology is the network
diameter which is greater than rest of the topologies 

\subsubsection*{DCell}
We implement 5 cell, 2 levels DCell topology for evaluation purposes.
Each cell has 4 hosts and an edge switch to connect to the servers
(hosts). In case of mininet, linux hosts have trouble forwarding
packets as they see each interface on the same network (broadcast domain).
An additional switch, therefore, was added as the edge switch
to ensure that the hosts could reach every destination.
In order to test all of the links in the topology simultaneously,
we place four hosts in each cell with a total of five cells. 
The limiting factor for the DCell topology is the single edge
switch in each cell. The single edge switch creates a bottleneck
in the cell limiting the performance of the topology.

\subsubsection*{Facebook Fat Tree}
We implement a smaller version of Facebook data center.
The smaller version has 48 edge routers and 4 aggregate routers.
The limiting factor for the Facebook design is the bottleneck
created by connecting 48 edge routers to the 4 aggregate routers.
When a 48 port switch or router has all of the connections
to other switches that have servers attached then it will be
tough for that switch to process packets as fast as they arrive.
Theoretically, the saturation throughput for a switch or routers
with N ports as N approaches infinity is 2 - $\sqrt{2}$
or around 58.6\%.

\subsection{Traffic Patterns}
We have evaluated the above topologies with a subset of following
traffic patterns -
\begin{itemize}
\item \textbf{\textit{Uniform Random Traffic:} }
Each packet is sent to any other node with equal probability.

\item \textbf{\textit{Bit Complement Traffic:} }
Each node exchanges packets with a node on the opposite side of
the network. To compute the destination address, a bit wise inversion
is carried out of the source coordinates. This traffic
provides a well balanced traffic across the network.

\item \textbf{\textit{Bit Reverse Traffic:} }
A message originating from a host having host address as $B_1 B_2 \ldots B_n$
is destined for a host with the address $B_n B_{n-1} \ldots B_1$.

\item \textbf{\textit{Tornado Traffic:} }
Each node $i$ sends traffic to $\left(i+ \frac{N-1}{2}\right) mod N$ where N is
total number of hosts.
\end{itemize}

\subsection{Mininet}

\input{mininet}

\subsection{Evaluation Using \textit{gem5}}
\input{gem5}

%% file: mininet.tex
For evaluation, we used Mininet as our network emulator
and POX as the OpenFlow controller.
The exact code can be found at
\url{https://github.com/lebiednik/ICNmininet}.
Mininet was chosen because of the ease of use with Python.
POX was chosen because it provided
the best network convergence rates among the OpenFlow controllers
tested (RIPL-POX, OpenDaylight, and Floodlight).  


Originally, we began programming the topologies with routers
instead of switches but with even a scaled version of the Google
Fat Tree topology there are forty networks or broadcast domains.
Thus, for each router (programmed as its own class of host in Mininet),
forty commands (network address and next hop) would need to be
entered before the router knows about every destination in
the network. Similarly,for DCell and Facebook topologies, the
routers would need to know about 35 and 240 networks respectively.

Tests were performed with \textit{iperf} and \textit{ping}.
The \textit{iperf} command can
calculate the bandwidth available between any two hosts by creating
a client-server Transmission Control Protocol (TCP) connection.
The ping command can not only provide connectivity results in
the network but can also be used for latency measures.
Ping also provides the capability to determine the size of the packet
that you are sending on the network which allows sending
larger packets to emulate sending larger files. All tests were
run on a dedicates 64-bit Ubuntu server with 8 vCPUs and 15 GB memory. 

\subsubsection*{Bisection Bandwidth Testing}
To test the bisection bandwidth, the team used iperf to create
Transmission Control Protocol (TCP) connections between bit
complement hosts. The iperf command provides a variable time
connection between two end devices, one will act as the host and
the other the server. The command provides a readout of the client
to server connection in bits per second (bps). Since iperf
acts as a client to server connection, the test would designate
one side as the client side and the other as the server side
and then switch the sides to pass as much data across the
bisection as possible. Each test was then averaged and
then provided a percentage of the total bisection bandwidth
to normalize the data.
\begin{figure}[h]
\centering
\includegraphics[scale=0.4]{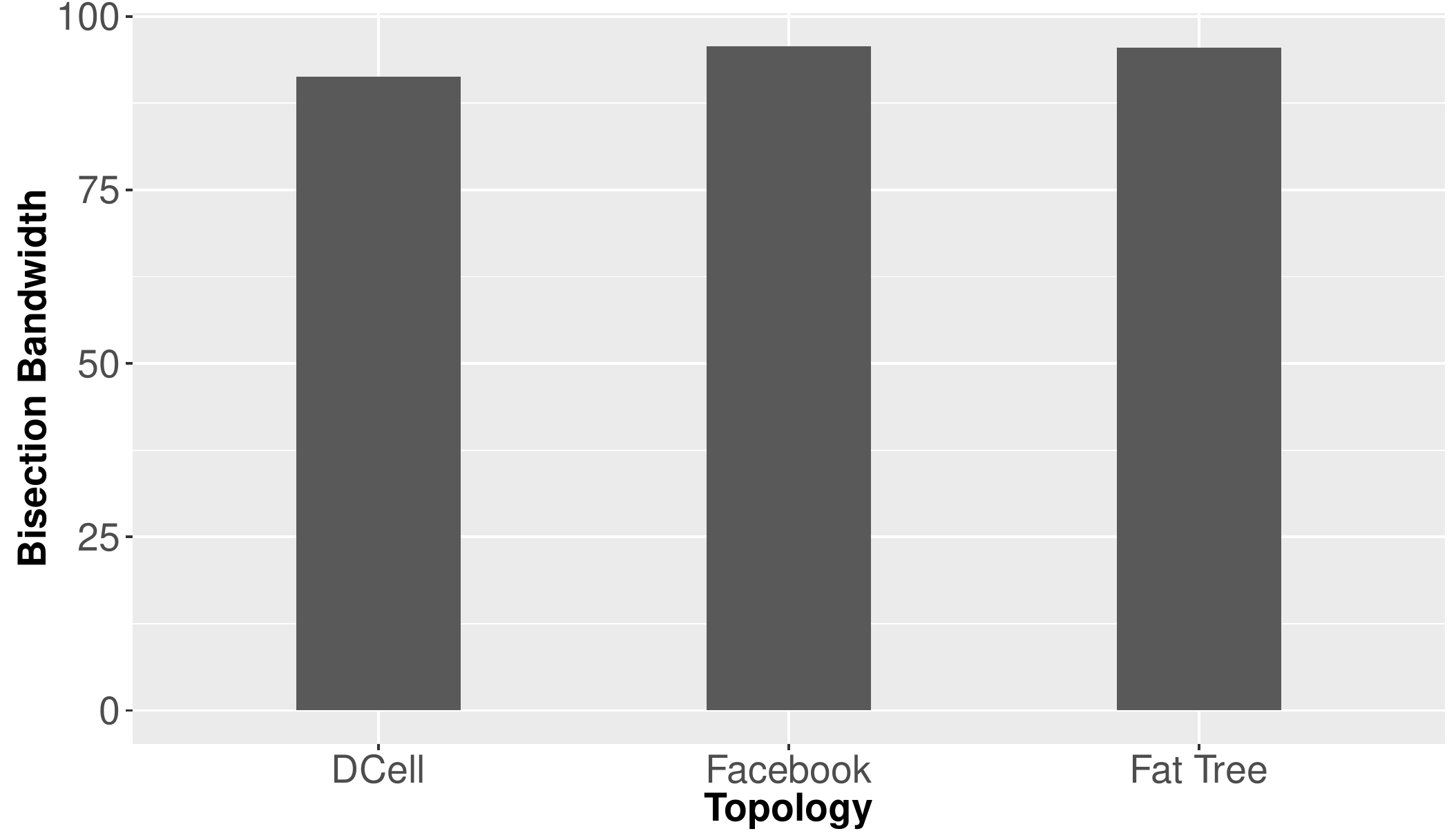}
\caption{Percentage of Bisection Bandwidth Achieved\label{fig:mininet_bisection_bw}}
\end{figure}

Figure \ref{fig:mininet_bisection_bw} shows that the Google Fat
Tree topology slightly outperforms the Facebook topology.
The test was run several times and the Google and Facebook
topology continually showed performance in the 95\% range
on all links. The DCell Topology provided around 95\%
utilization on most links but with the bottleneck (many connections to the same router) close
to the hosts, one link would show around 78\% total utilization
dropping the average utilization greatly. Since DCell is
using an edge switch or the multihoned (multiple connections to the same host) server, this result
would make sense as the server would have greater difficulty
multiplexing on all of the links simultaneously while
an edge switch would create a bottleneck in the network. 

\subsubsection*{Packet Size Testing}
The Maximum Transmission Unit (MTU) is the largest packet
allowed on the network links. Based off old Ethernet standards,
the largest packet allowed on the network is 1500 bytes.
Routers and hosts drop anything larger to prevent "Ping of Death"
attacks and to maintain the standard. For comparison, the normal
ping packet size is 82 bytes. Network designers would
know the average packet size that they would want to
design for their networks. For example, moving large
amounts of data within a data center would require lower latency
with larger packets. The design would need to prevent bottlenecks.
Whereas a network that has a smaller average packet size could
have potential bottlenecks as long as the link speeds were
fast enough.
\begin{figure}[h]
\centering
\includegraphics[scale=0.4]{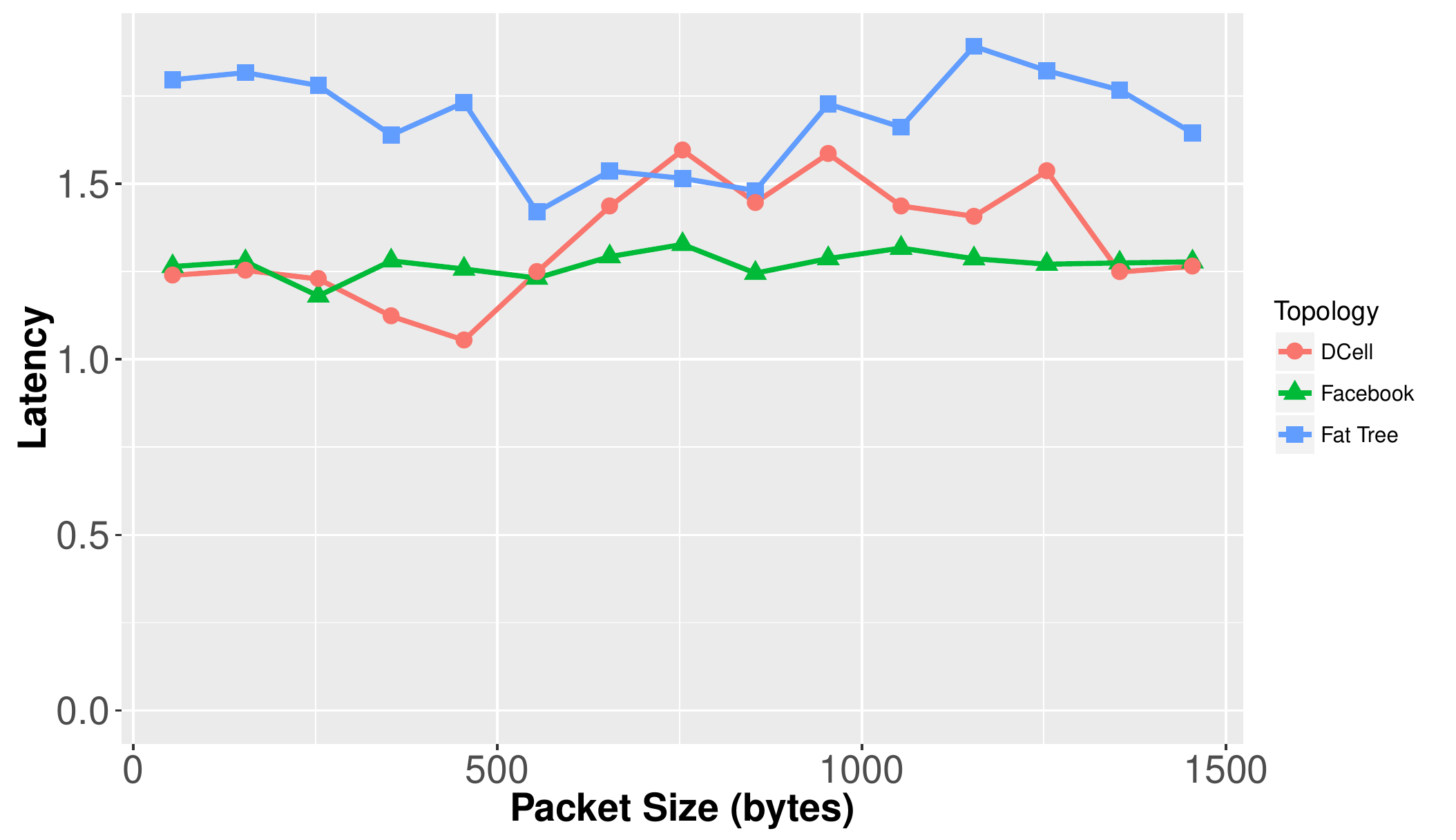}
\caption{Average Latency (ms) vs Packet Size(bytes)\label{fig:mininet_latency}}
\end{figure}

Figure \ref{fig:mininet_latency} shows the average latency of each
of the networks versus the size of the packet placed on
the network. As expected the Facebook topology provides
the lowest latency in the network because of the diameter
of the network. The DCell topology begins to bottleneck
when the size of the packet increases over 200 bytes.
The DCell Topology still provides less latency.
The Facebook topology displays the lowest latency
despite the size of the network and the potential
for bottlenecks. The Facebook topology has a shorter
overall diameter but the aggregation of 48 edge
routers with hosts to four aggregate routers has
the potential for a large bottleneck when compared
with the Google Fat Tree which has equal links between
all of the devices in the topology.  

\subsubsection*{Network Saturation Testing}
To test the networks during periods of high utilization,
hosts pinged their bit complement with a maximum packet
size for a three times as long as the other tests.
The purpose of this was two-fold. First, it would provide a
greater amount of time for the network to reach saturation,
much like test in NoCs perform. Second, the team wanted
to test the bisection capacity with the largest packets
sizes available on the network.

\begin{figure}[h]
\centering
\includegraphics[scale=0.35]{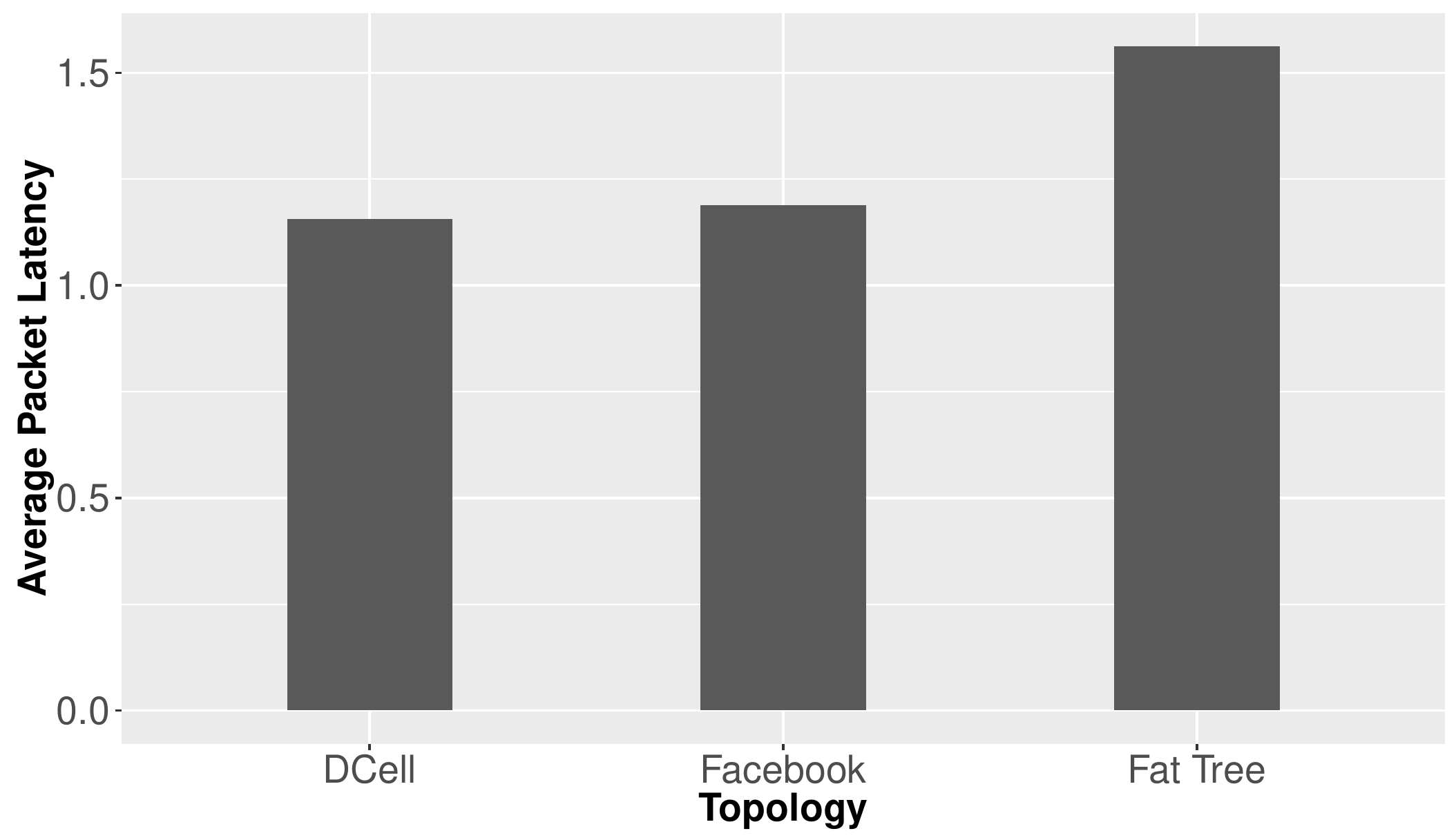}
\caption{Latency Test for 30 seconds at MTU(ms)\label{fig:mininet_stress_latency}}
\end{figure}
Figure \ref{fig:mininet_stress_latency} shows the average latency
for packets sent in over the course of the 30 second test.
The DCell topology slightly outperforms the Facebook
topology by less than a tenth of a millisecond. Based
on the diameter of the network and the hops each packet
would take, Facebook should outperform DCell. This test
could show that Facebook reaches its saturation and starts
under-performing due to the bottleneck at the core level.
The bottleneck of the Facebook topology is much bigger than
the bottleneck of the DCell with an average of 12
connections from the edge level to the aggregate level.
Whereas, DCell has 4 hosts connected to one router at the
edge level. The diameter of the Google Fat Tree topology
causes the higher latency in the network. Based off these
results, the topology diameter has a significant impact
on the latency but so does any bottlenecks in the network.

\begin{figure}[h]
\centering
\includegraphics[scale=0.35]{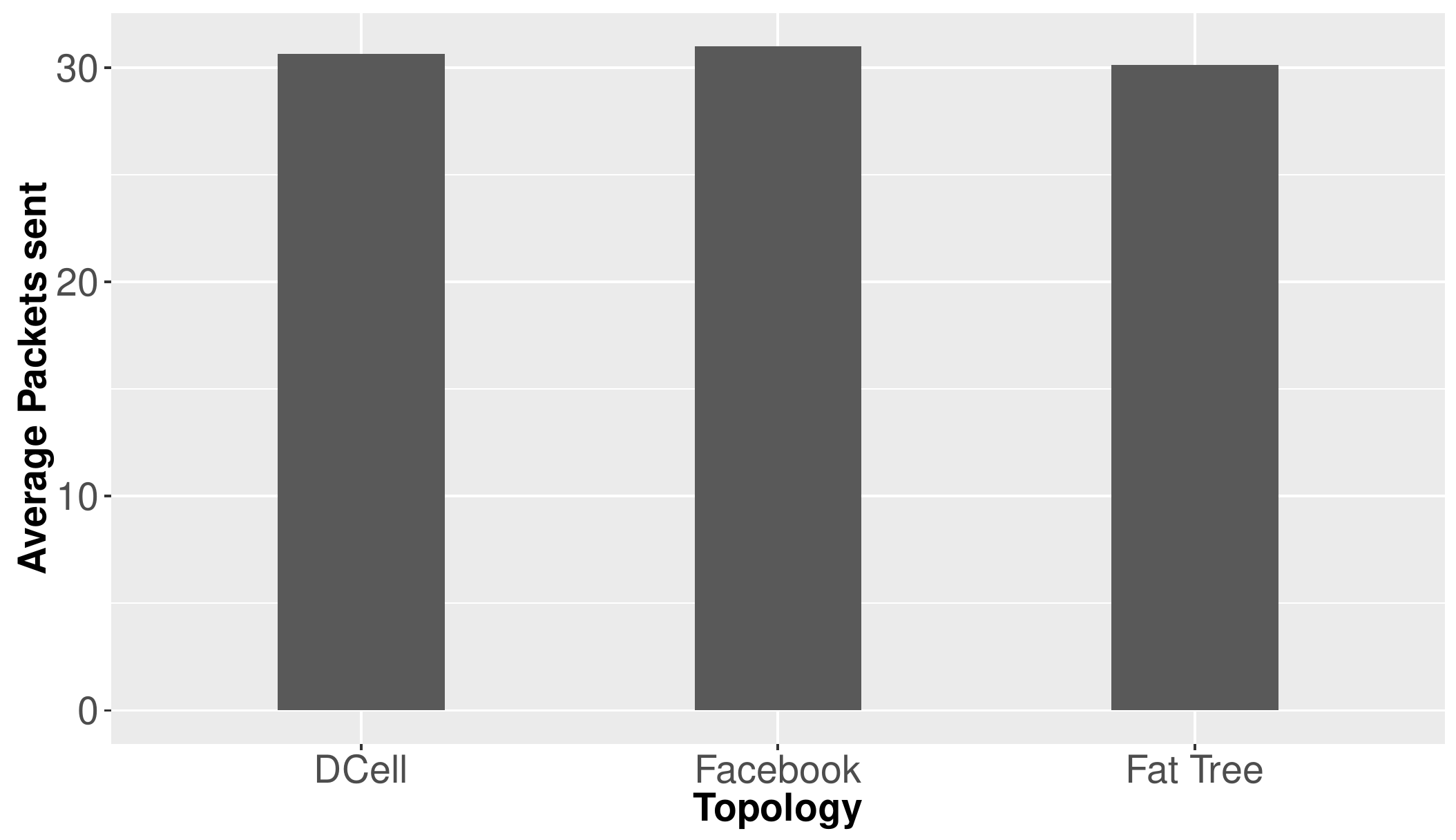}
\caption{Packets Sent During Stress Test per host\label{fig:mininet_packets_sent}}
\end{figure}
Overall, during the stress test, each host was able to send over
30 packets apiece successfully. Figure \ref{fig:mininet_packets_sent}
provides the average number of packets sent by each host in
each topology. The Facebook hosts sent almost one packet more
per host during the period than the Google Fat Tree.
So in comparison, the Facebook topology was able to send
48 more packets or 72KB with the 10 Mbps links. 

\subsubsection*{Scalability Testing}
To further evaluate the network performance for each of
the topologies, the team began adding more hosts to the edge
switches. This scalability testing is important as it
helps to determine the optimal number of servers that the
data center engineer would want to place under each edge switch.

\begin{figure}[h]
\centering
\includegraphics[scale=0.4]{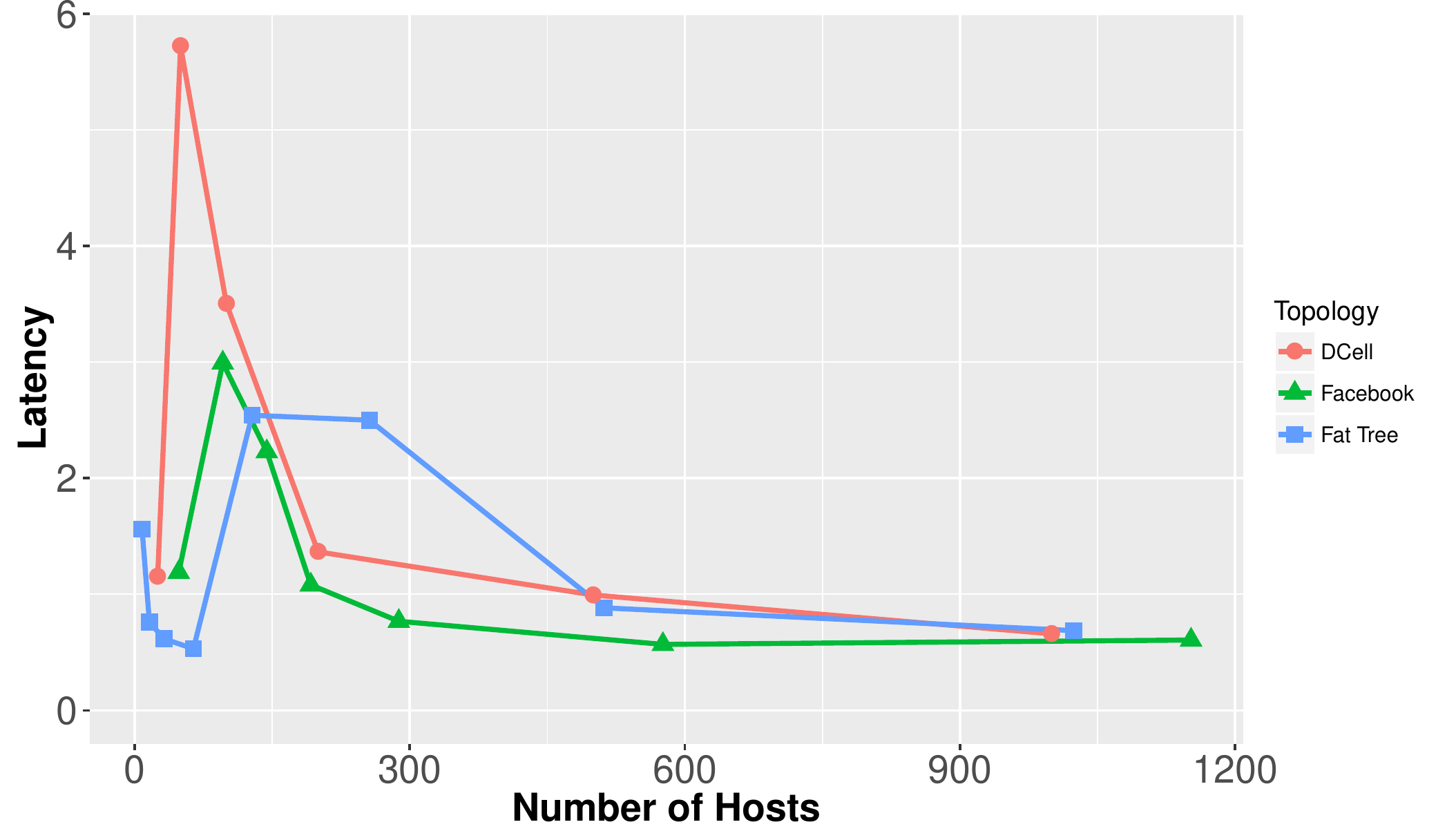}
\caption{Traffic Latency versus number of hosts(ms)\label{fig:mininet_scaling}}
\end{figure}
As shown in Figure \ref{fig:mininet_scaling}, each topology provides
less latency at different amounts of hosts.
The team ran the simulations several times at each
of the points and obtained similar results. Although
at the bases state Fat Tree has greater latency, by the
time that there are a similar amount of hosts on the network,
Google Fat Tree has less latency than the Facebook topology.
All of the topologies have an interesting peak latency in
the 50 to 300 host range but then level off towards an equilibrium.
The Facebook topology outperforms all others from 144 hosts onward.

%% file: gem5.tex
\textit{gem5} \cite{binkert2011gem5} is a simulation platform for
computer-system architecture research. It came as a merger
of the \textit{m5} simulator \cite{binkert2006m5} from the
University of Michigan Ann Arbor and the \textit{GEMS} simulator
\cite{martin2005multifacet} from the University of Wisconsin Madison.
\textit{gem5} uses \textit{garnet} to simulate the
interconnection network for NoCs.
We modeled Google Fat Tree and DCell topology on
Garnet network simulator \cite{agarwal2009garnet}
by modifying it to emulate the working
of an actual off chip network, as closely as possible. 

First, the router delay is increased to five cycles
by increasing the number of stages in the router pipeline.
The link latency was increased to 10 cycles to model higher
link propagation latency.
We used table based routing, which is already available in Garnet.
We further modified the algorithm to choose one of the
redundant paths randomly.
For example, in Google Fat Tree topology, a packet has more than one
possible path while going up the tree. However, once the packet
reaches the core router, it has only one available path going down
the tree. We choose a path randomly while going up the tree to achieve
higher link utilization.
An important difference is that off chip networks have a much higher
number of queues than on chip networks, and so we simulate
that by having number of virtual channels as equal to 100.
We also consider one flit per packet, to simplify our assumptions.
In this work, We attempt to implement dropping of packets as
is done commonly in data center networks. Packets are randomly
dropped and re transmitted by the sender if an available
outgoing queue is not available. We plan to compare the
results of this implementation with having the packets wait
for their turn, to see latency benefit in each, as the
future work of this project.
We ran several simulations on Google Fat Tree topology
and Microsoft's DCell topology, the results of which have
been discussed below.   

\subsubsection*{Google Fat Tree: Throughput across different traffic patterns: }
\begin{figure}[h]
\centering
\includegraphics[scale=0.36]{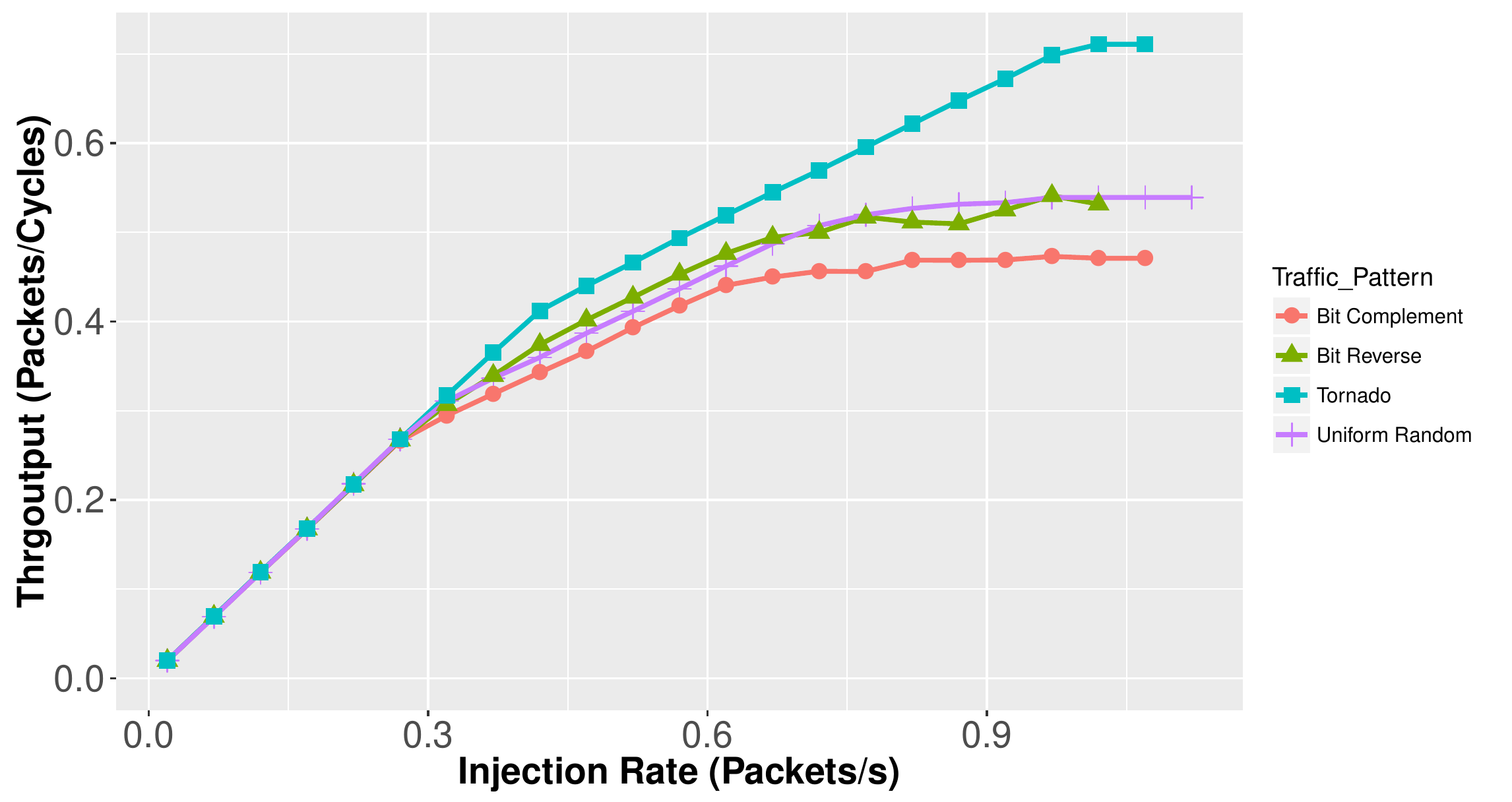}
\caption{Google Fat Tree: Throughput across different traffic patterns\label{fig:gem5_traffic_throughput}}
\end{figure}

Figure \ref{fig:gem5_traffic_throughput} shows the values of packet
reception rate (total\_packets\_received/num-cpus/sim-cycles)
at different injection rates for four different traffic patterns - uniform
random, bit reverse, bit complement, and tornado. It can be seen in
this figure that, almost all traffic patterns except tornado show
similar reception rate. The reason for this is, in the Google Fat
Tree topology, if the destination host is outside of the pod
belonging to the source, it will take the same number of hops,
for all cases, except tornado. Hence, all patterns present
results similar to that of uniform random pattern. However,
in the case of tornado, we see a much better reception rate
for the reason being that due to nature of tornado traffic,
it is possible that some of the source and destination pairs
are within the same pod, which results in a much higher reception rate.

\subsubsection*{Google Fat Tree: Packet Reception Rate for different
number of virtual channels:} 
\begin{figure}[h]
\centering
\includegraphics[scale=0.38]{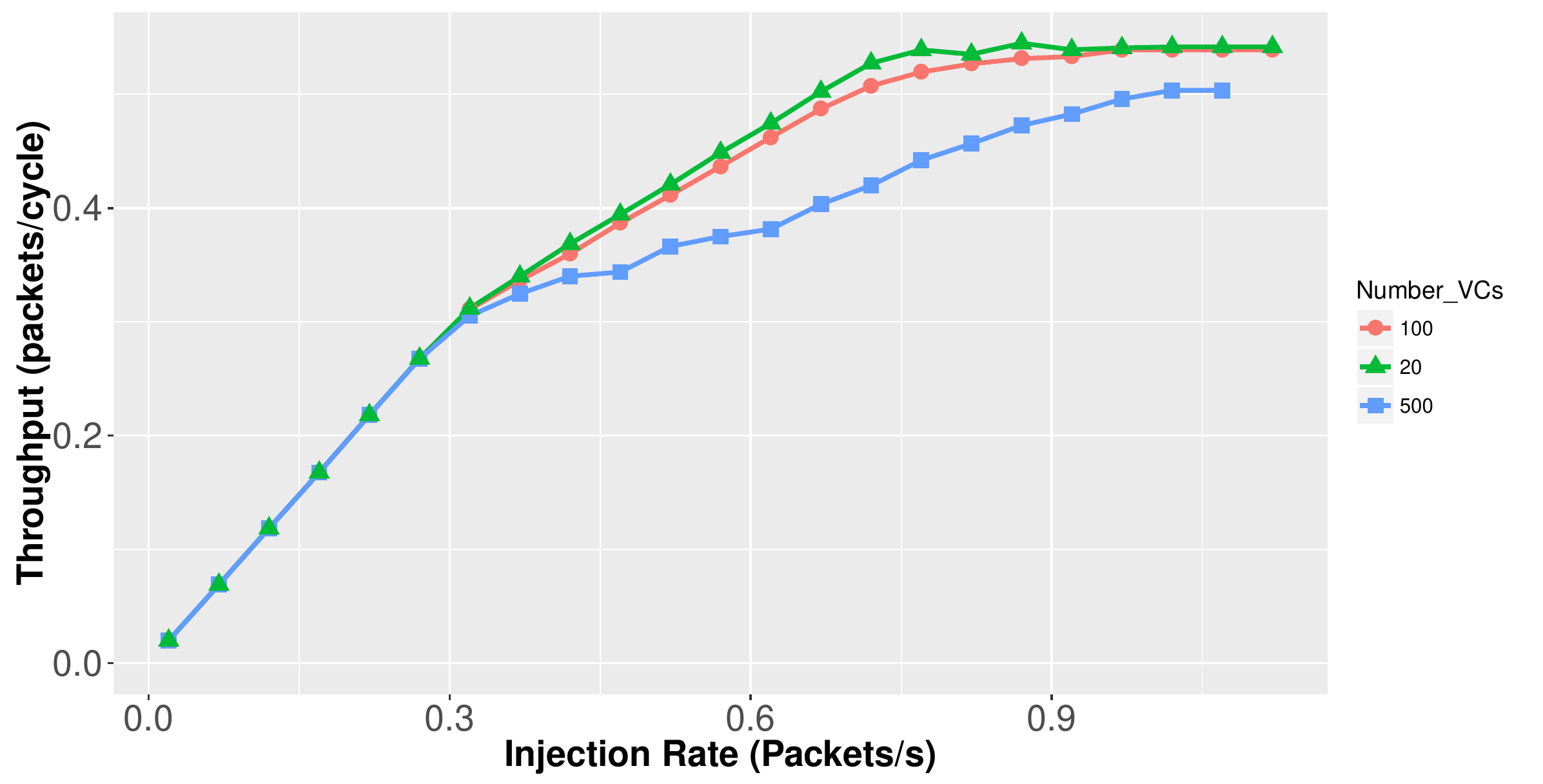}
\caption{Google Fat Tree: Packet Reception Rate for different number of virtual channels\label{fig:gem5_vc}}
\end{figure}

Figure \ref{fig:gem5_vc} shows
the packet reception rate for Google Fat tree topology across
three different number of virtual channels - 20, 50 and 100.
The packet reception rate does not necessarily get better
with more number of virtual channels, the reason for this
being, there is no contention delay to begin with, due
to having multiple number of VCs.

\subsubsection*{DCell: Throughput across different traffic patterns:}
\begin{figure}[h]
\centering
\includegraphics[scale=0.36]{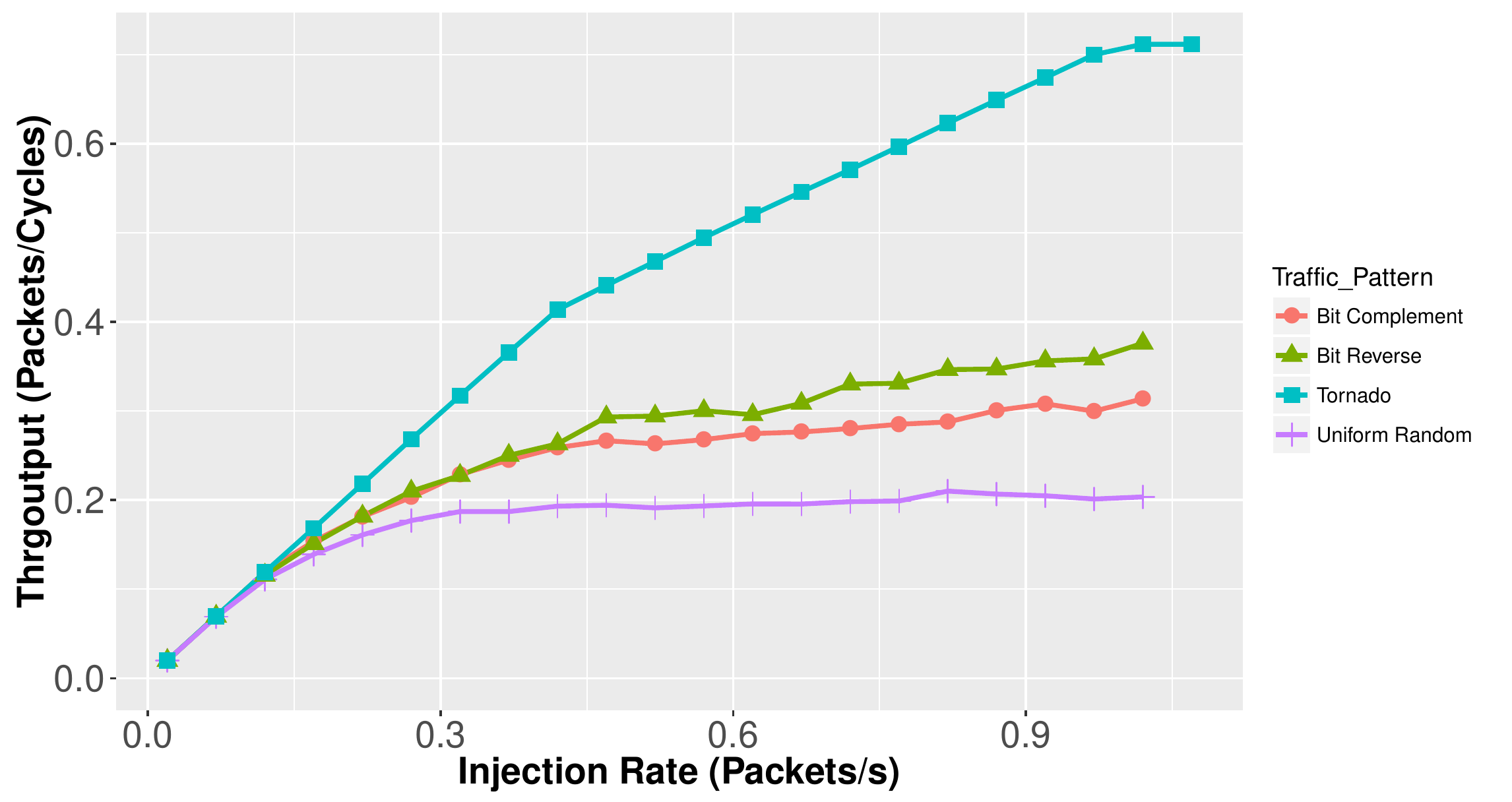}
\caption{DCell: Throughput across different traffic patterns\label{fig:gem5_dcell_traffic_throughput}}
\end{figure}

Figure \ref{fig:gem5_dcell_traffic_throughput} hows the
throughput values obtained for DCell topology over four
different traffic patterns. Again, Tornado shows much
better throughput as compared to other traffic patterns.
The reason for this is that tornado traffic pattern does
a much better job of distributing the traffic among the
links in the network, whereas the other traffic patterns,
tend to strain few links more than other.

\subsubsection*{Throughput Comparison of Google Fat Tree and DCell}
\begin{figure}[h]
\centering
\includegraphics[scale=0.36]{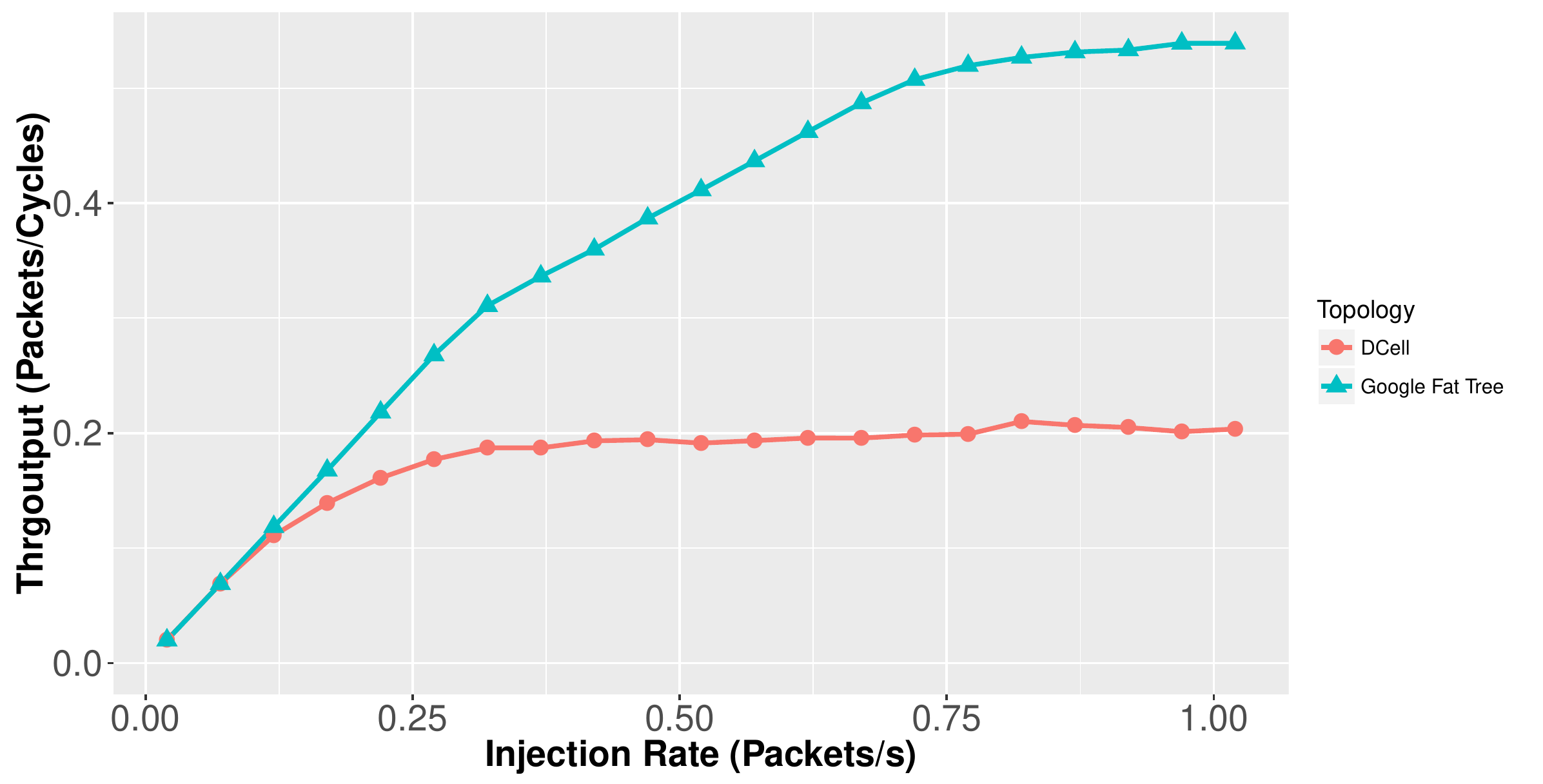}
\caption{Throughput Comparison of Google Fat Tree and DCell\label{fig:gem5_dcell_google}}
\end{figure}

Figure \ref{fig:gem5_dcell_google} shows the throughput values
of Google topology to be much higher than that of DCell.
This is as expected because the Google topology has a much
higher ratio of number of links to number of hosts. DCell
essentially has one switch per module, no matter how many
servers are there per module, and so that causes a bottleneck
in DCell performance. 


\subsubsection*{Throughput Comparison of DCell for 20 hosts and 42 hosts}
\begin{figure}[h]
\centering
\includegraphics[scale=0.36]{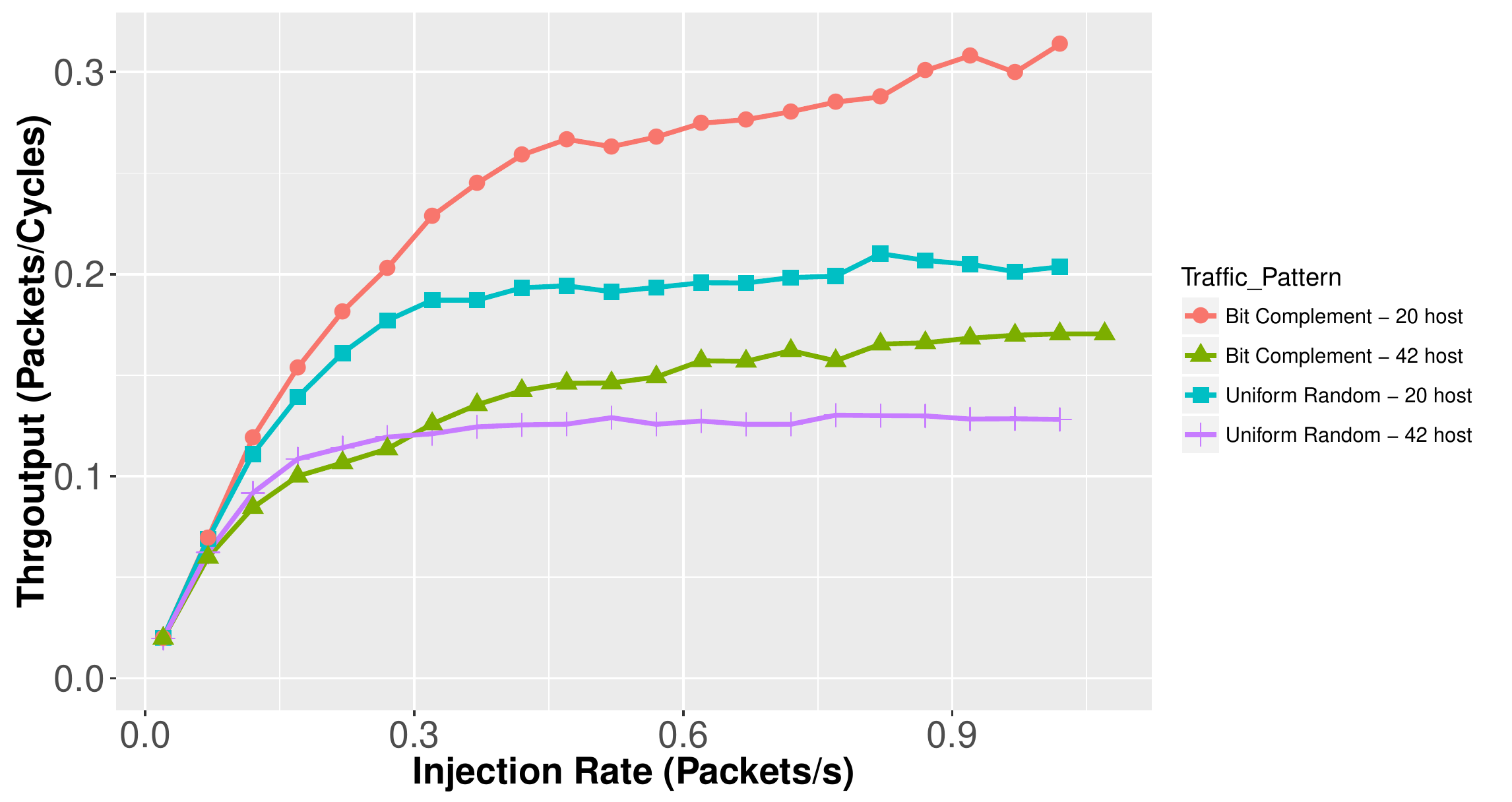}
\caption{Throughput Comparison of DCell for 20 hosts and 42 hosts\label{fig:gem5_dcell_throughput_hosts}}
\end{figure}

Figure \ref{fig:gem5_dcell_throughput_hosts} shows throughput
values for two different values of n. For n=4, the number of
hosts are 20, and for n=6, the number of hosts are 42.
We simulate the throughput for both topologies across
two different traffic patterns - uniform random, and
bit complement traffic. The saturation throughput reduces,
as we increase number of hosts, because, as discussed above,
the number of servers per module switch increases, essentially
overloading the switch. This results in a lower throughput
as we increase number of hosts. This result is consistent
with the conclusion that DCell is essentially a
recursive topology, which scales by increasing the
number of levels in the topology, as opposed to increasing
the number of servers within one level.
This results serves to prove this fact. 

\subsubsection*{DCell: Latency comparison across various different traffic patterns}
\begin{figure}[h]
\centering
\includegraphics[scale=0.36]{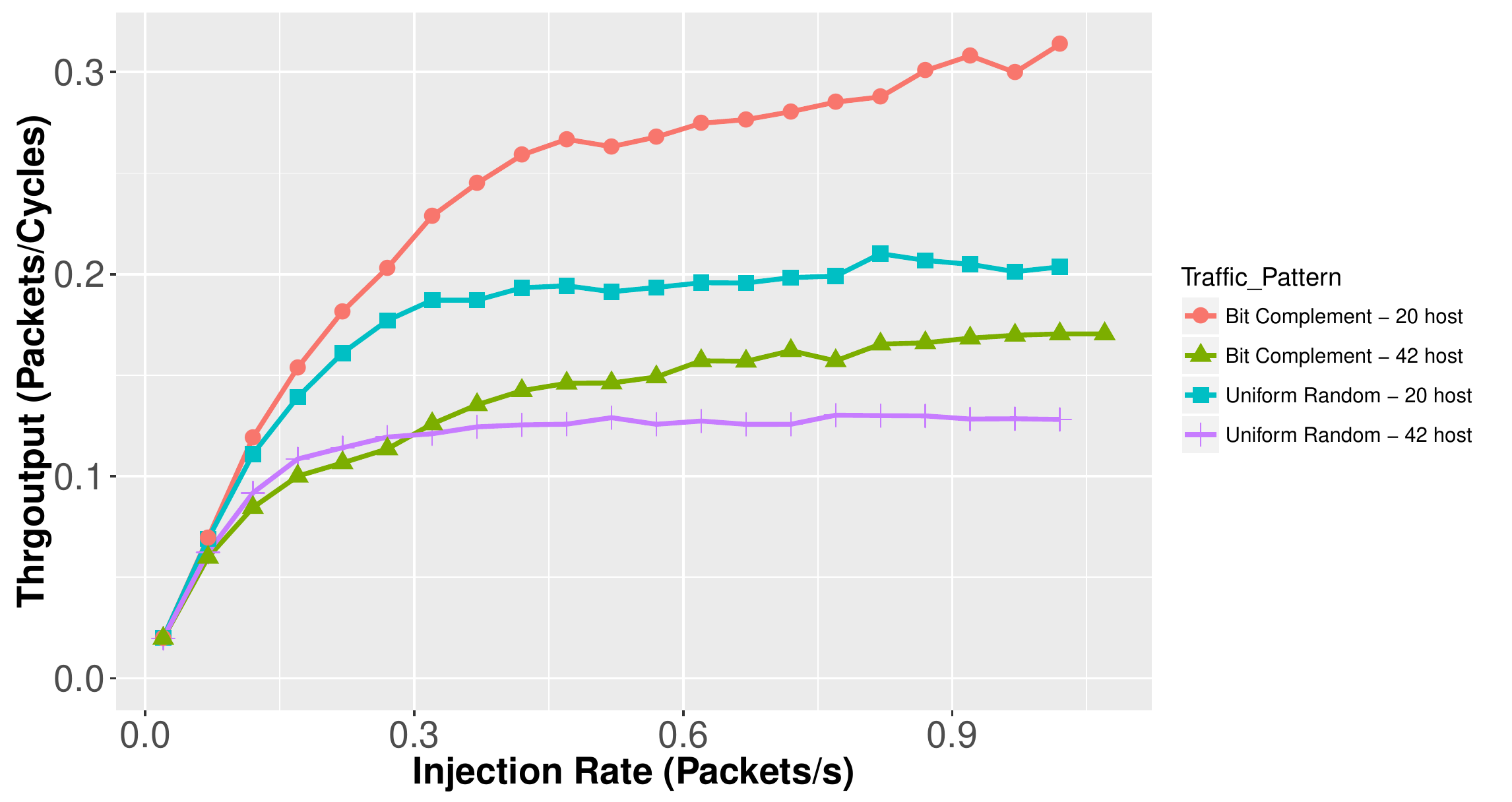}
\caption{DCell: Latency comparison accross various different traffic patterns\label{fig:gem5_traffic_latency}}
\end{figure}

Figure \ref{fig:gem5_traffic_latency} shows the latency comparison
between different traffic patterns of DCell.
As discussed in the throughput comparison, tornado better
distributes the traffic across the network, and hence,
presents better latency than the other traffic patters. 

%% file: conclusion.tex
By testing the Google Fat Tree and DCell topologies in both gem5 and Mininet, we were able to produce similar results. This helps to verify the results of testing multiple topologies. Both testing in Mininet and gem5 showed that the Google Fat Tree topology outperforms the DCell topology in throughput.

While our research produced some interesting results in regards to the topologies, there is certainly more work to be done. For our Mininet simulations, we would like to build a controller that is not only able to function with network loops but also provides load balancing. Currently, only the controllers used in production environments, such as Google, are able to control networks with loops and provide load balancing. Most other controllers that work with loops simply use the Spanning Tree Protocol to shut off links that may be redundant.

Implementing the off-chip topologies in gem5 simulator, gave us a lot of insight into the working of the internconnection network in an off chip topology. We have successfully implemented and simulated several of the parameters of data center networks, on garnet. The next steps with regards to implementation in garnet, could be the consideration of scalability of DCN as well as the implementation of packet dropping in garnet, which is a common phenomenon in data center topologies.

During the scaling of hosts testing, Mininet provided interesting results. The team has not previously seen Mininet topologies with such large amount of hosts. Further investigation is required to understand why latency in the network increase drastically for up to 300 hosts and then decreases. The results from the Amazon EC2 server were consistent even when the tests were run several times. While Facebook provided the best overall latency, there are several instances where the other topologies outperform it in regards to the same or approximately the same number of hosts.

There are other traffic patterns that other studies have used to evaluate their designs. In one of the papers, the authors used Stride traffic patterns to mimic their datacenter traffic. This would be another good comparison for the datacenter topologies. We used bit complement for most of the tests in Mininet because it had the same hop count as it would have with the unaltered datacenter.

We have presented a comprehensive survey and taxonomy of
Data Center Network topologies that have been proposed
in the history of off chip networks.
Even though a significant number of topologies
have been explored, only a few such as Fat Tree, BCube
have been implemented in practice in order to achieve
high bisection bandwidth. Today, the principle bottleneck in
large scale cluster is often inter-node communication
bandwidth. To keep up with this demand, we need scalable
network topologies that can fulfill significantly high
bandwidth requirements while keeping the cost low.

We have also presented a comparison of a few data center
topologies using \textit{mininet} and \textit{Gem5} simulator.
We plan to complete our analysis as part of (short term)
as part of our future work.